\documentclass[11pt]{article}

\usepackage{arxiv}

\usepackage[utf8]{inputenc} 
\usepackage[T1]{fontenc}    
\usepackage{hyperref}       
\usepackage{url}            
\usepackage{booktabs}       
\usepackage{amsfonts}       
\usepackage{nicefrac}       
\usepackage{microtype}      
\usepackage{lipsum}
\usepackage{graphicx}
\graphicspath{ {./images/} }

\usepackage{amsmath}
\usepackage{natbib}
\usepackage{authblk}
\usepackage{siunitx}
\usepackage{moreverb}
\usepackage{mathtools, cuted}
\usepackage{rotating}
\usepackage{appendix}
\title{Information transfer as a new indicator of approaching local bifurcations: saddle-node, pitchfork and Hopf bifurcations}
\date{}

\author[1]{St\'ephane Vannitsem}
\author[2]{Carlos A. Pires}
\author[3,4]{X. San Liang}

\affil[1]{Nanyang Technological University, Singapore}
\affil[2]{Instituto Dom Luiz, University of Lisboa, Portugal}
\affil[3]{Department of Atmospheric \& Oceanic Sciences, Fudan University, China}
\affil[4]{The Artificial Intelligence Department, Division of Frontier Research, Southern Marine Science \& Engineering Guangdong Laboratory (Zhuhai), China}

\begin{document}

\maketitle

\begin{abstract}
Early warning signals are key tools for anticipating qualitative changes in the dynamics of complex systems. In this work, we investigate whether an information-theoretic measure designed to characterize directional causal dependencies between observables can provide early warning of the approach to bifurcations. The measure considered is the rate of information transfer, which has been successfully applied to characterize directional dependencies in a broad range of fields, from economics to climate science. Three fundamental local bifurcations are considered: saddle-node, pitchfork, and Hopf bifurcations. We first investigate idealized (quasi-)potential stochastic systems for which analytical expressions for the rate of information transfer can be derived. Distinct signatures are found for the three bifurcation types. In particular, the rate of information transfer displays a strong increase and divergence when approaching the saddle-node bifurcation, whereas smoother but systematic changes are found near the pitchfork and Hopf bifurcations. The approach is then tested using low-order climate models exhibiting the same types of bifurcations. The results confirm that changes in the rate of information transfer can precede the corresponding dynamical transitions, suggesting that directional information transfer provides a useful complementary early-warning indicator for detecting the approach to critical transitions in nonlinear stochastic systems.
\end{abstract}

\section{Introduction}

Early warning signals of catastrophic events have always been searched for. One particularly relevant catastrophic event in physical sciences is the approach of saddle-node or limit point bifurcation \citep{Pomeau2011, Peters2012, Kuehn2013}, which is believed to be a highly probable scenario of collapse of important dynamical processes of the climate system like the Atlantic Meridional Overturning Circulation (AMOC), or in ecosystems, as discussed for instance in \citep{Scheffer2009,Boulton2014,Boers2021,Boers2021b,Ditlevsen2023,Wunderling2024}. In recent years, methods have been developed and proposed based primarily on the autocorrelation and probability distribution of the signal embedded in a time series, but also on more sophisticated concepts coming in particular from dynamical systems theory, see for a nice overview \citep{Liu2026}. In the present work, we will explore the possibility to construct early warning indicators of approaching bifurcations using recent causal techniques of information theory involving several joint variables. The hypothesis behind this exploration is that besides the potential changes of variability within a time series close to a bifurcation or a tipping point, there should be some related modifications of the relationships between the set of joint variables. This hypothesis is at the core of our analysis, and as it will be shown, there are indeed drastic changes that can help providing early warning signals of the approach of bifurcations.

The concept of rate of information transfer between variables has been introduced in 2005 by \cite{Liang2005, Liang2007}. It is based on the rate of change of the marginal Shannon entropy of a specific variable, and decomposed into a self-influence term and terms related to the transfer of information of the other variables of the system, see \cite{Liang2021} for a recent overview. This approach has been very successful in several applications thanks to the development of the theory for linear stochastic systems that could directly be applied on time series \citep{Liang2014,Stips2016,Vannitsem2019,Hagan2019,Docquier2022,Hagan2022,Vannitsem2022,Docquier2023,Docquier2024,Siddique2026}. The linearity assumption although very useful as a primary tool, is a limitation when dealing with nonlinear systems as discussed in details in \cite{Pires2024,Vannitsem2024,Vannitsem2025}. The paper of \cite{Pires2024} offers a very interesting way to compute the rate of information transfer in multivariate nonlinear systems by computing conditional expectations, instead of relying on the explicit estimation of the joint probability or conditional densities. The relations deduced in this paper will also be extensively used in the current work.

In the present work, we will explore the change of properties of the rate of information transfer close to bifurcations (saddle-node, Pitchfork and Hopf) that are commonly found in different fields \citep{Guckenheimer1983,nicolisbook2}, and in particular in atmospheric and climate sciences. We will first investigate analytically and numerically three potential or quasi-potential stochastic 2-dimensional systems in which such bifurcations occur. We will then turn to a set of systems that describe the dynamics of important climate processes, namely the thermohaline circulation \citep{Stommel1961, Cessi1994}, the gyre dynamics in the upper layer of the ocean \citep{Jiang1995} and the recharge-discharge oscillator describing the large-scale dynamics of the ocean in the tropical Pacific \citep{Jin1997}. Stochastic noise are added to those three 2-dimensional systems to generate variability. As it turns out, the rate of information transfer in its nonlinear form constitutes an interesting indicator of the proximity to the bifurcations, while the linear approach can sometimes indicate changes but not in a systematic manner. 

Section 2 briefly presents the formalism of the rate of information transfer. In Section 3, the analytical computation of the rate of information transfer in the potential or quasi-potential systems is performed, together with the numerical validation of the analytical results. Section 4 is then presenting the analysis of the three climate-relevant systems, and Section 5 summarizes the main conclusions.

\section{The formalism for information transfer}

The development of the formalism on the rate of information transfer has a long history, starting from the work of \cite{Liang2005}. Recently the theory has been revisited to be able to compute easily the quantities involved without computing explicitly the full joint probability of the set of variables under interest \citep{Pires2024}. Let us consider a dynamical system written as
\begin{equation}
 \frac{d\bf{{X}}}{dt} = {\bf F} ( {\bf X} , {\bf \lambda}) + {\bf B} \frac{d{\bf W}}{dt}
 \label{dynsystem}
\end{equation}
where ${\bf X} = (X_1, ..., X_N)$ with $N$ the number of variables, and $\bf{\lambda}$, a set of parameters. No explicit dependence on time is appearing in the right hand side of the equations. The system is also forced by a noise term where ${\bf W}$ is a vector of Wiener processes of dimension $N$. The matrix ${\bf B}$ is the covariance matrix of the noise, which will be considered as diagonal and not dependent on the variables themselves in the following. This considerable simplifies the estimation of the rate of information transfer between variables, which is then limited to the relations below.

The formalism of \cite{Pires2024} starts from the rewriting of the rate of information transfer as
\begin{eqnarray}
T(X_{\neq i} \rightarrow X_i) & = & E \left[  F_{i} \frac{\partial log(p({ X_{\neq i}|X_i}))}{\partial X_i}  \right] \nonumber \\ & = & E \left[  F_{i, \neq i} \frac{\partial log(p({X_{\neq i}|X_i}))}{\partial X_i}  \right] 
\label{RIT}
\end{eqnarray}
where $F_{i, \neq i}$ is the part of the deterministic term $F_{i}$ which contains all the terms influencing $X_i$ that are not exclusively dependent on $X_i$ and $p(X_{\neq i}|X_i)$ is the conditional probability density function of all the variables $X_k, k \neq i$ given $X_i$. $E[.]$ denotes the expectation over all variables.

\cite{Pires2024} showed also that the rate of information transfer can be further expressed in terms of the conditional expectations of the variables when $F_{i, {\neq i}}$ is a sum of separable functions like multivariate polynomial functions,

\begin{equation}
T(X_{\neq i} \rightarrow X_i) =  \sum_{l} E_i \left[  f_{1,l} (X_i) \frac{\partial E_{k \neq i} (f_{2,l}(X_{ k \neq i})| X_i)}{\partial X_i}  \right]
\label{totalret}
\end{equation}
where $F_{i, {\neq i}}$ is written as $ \sum_{l} f_{1,l} (X_i) f_{2,l}(X_{k \neq i})$. 

In the previous formulation, the dependency was formulated by considering the whole set of variables that are different from the target variable $X_i$. This can be further decomposed into contributions from each variable and their synergetic terms. To have more information on those decompositions, we refer to \cite{Pires2024}. As we will be here dealing with systems of 2 variables only, the above expression is exact for the dependence from one variable, say $x$, to another variable, say $y$. We will use these new notations in the following.

The two Equations \ref{RIT} and \ref{totalret} will be central in the analysis that will be performed below close to bifurcation. The first one (Eq. \ref{RIT}) will be used directly on the Mexican Hat potential model to compute explicitly the rate of information transfer from the probability densities of the system, while the second (Eq. \ref{totalret}) will be used for the numerical verification of the first result. This second expression will also be used on the Stommel-Cessi model of the North Atlantic overturning circulation.

Finally it is also useful to evaluate the impact of the use of simplified approximations of the rate of information transfer. An important simplification that can be made when the underlying dynamics is not known, is the linear assumption. In this case the rate of information transfer can be computed straightforwardly for bivariate or multivariate set of variables \cite{Liang2014, Liang2021}. As we are dealing in the current setting to bivariate systems, we will use the estimator that was deduced in \cite{Liang2014}, It has been shown that 
that the maximum likelihood estimator (MLE)
of the information transfer from any variable $x$ to variable $y$, 
	\begin{eqnarray}	\label{eq:T21_hat_2d}
	\hat T_{x \to y} = \frac {C_{xx}C_{xy} C_{y,dx/dt} - C_{xy}^2 C_{x,dx/dt}} 
			  {C_{xx}^2 C_{yy} - C_{xx}C_{xy}^2}.
	\end{eqnarray}
where the entries of the covariance matrix $ {\bf C} =
(C_{ij})$ are appearing, together with the covariances between the tendency and the variables themselves, $C_{x,dx/dt}$ and $C_{y,dx/dt}$. The indices can be inverted to get the opposite transfer of information. A clear asymmetric structure is found, and when covariance is 0, the tranfer of information is also 0 (nil causality principle). This method will also be used in the current work.

Note that in the following the notation introduced at the beginning of the section will be simplified as we consider only 2 variables, $x$ and $y$. 

\section{Idealized systems close to bifurcation}

An important category of systems for which the computation of the rate of information transfer (RIT) is easier is the one for which the deterministic part of the dynamics is potential. It allows to get access to the analytical expression of the probability densities. We will therefore consider in this section such systems in two dimensions, 
\begin{align}
\frac{dx}{dt}&=-\frac{\partial U}{\partial x}+\sqrt{2D}\,\frac{dW_x}{dt},\\
\frac{dy}{dt}&=-\frac{\partial U}{\partial y}+\sqrt{2D}\,\frac{dW_y}{dt},
\end{align}
where $U(x,y)$ is the potential, $W_x$ and $W_y$ are Wiener processes. The variance of the noises is equal to $2 D$. Two different potentials will be considered, leading to different local bifurcations, a saddle-node bifurcation and a Pitchfork bifurcation, given respectively by
\begin{eqnarray}
U(x,y) & = & \frac{\alpha}{4}(x^2+y^2-1)^2-\varepsilon x \\
U(x,y) & = & -\frac{\mu}{2} x^2 + \frac{x^4}{8} + \frac{1}{2} (y+\frac{x^2}{2})^2
\end{eqnarray}

The first one is known as the potential of the Mexican Hat. The two system of equations are therefore for the saddle-node bifurcation,
\begin{eqnarray}
\frac{dx}{dt} &=& -\alpha\,(x^2+y^2-1)x+\varepsilon+\sqrt{2D}\,\frac{dW_x}{dt}, \nonumber \\
\frac{dy}{dt} &=& -\alpha\,(x^2+y^2-1)y+\sqrt{2D}\,\frac{dW_y}{dt}.
\label{Eq:Saddle-node}
\end{eqnarray}
 The stationary solutions are provided in Appendix \ref{Appendix1}. Note that for $\varepsilon$ different from 0, the only possible solution is for $y_s=0$. A limit point bifurcation is found for the variable $x_s$. In the following the information transfer will then be deduced around the stable branch emerging from the saddle-node for negative values of $\alpha$ and a positive value for $\epsilon$ fixed here at $0.2$. Note that since the potential is not globally confining for negative values of \(\alpha\), all normalization integrals and statistical expectations are evaluated over a bounded domain \(\Omega\) containing the stable equilibrium. This restriction provides a locally normalized approximation in the weak-noise regime. Boundary contributions to the information-transfer expressions are neglected when the potential barrier separating the equilibrium from the domain boundary is sufficiently large compared with the noise intensity.

For the Pitchfork bifurcation, the equations are
\begin{eqnarray}
\frac{dx}{dt} &=& \mu x - x^3 -x y +\sqrt{2D}\,\frac{dW_x}{dt}, \nonumber \\
\frac{dy}{dt} &=& -y - \frac{1}{2} x^2 +\sqrt{2D}\,\frac{dW_y}{dt}.
\label{Eq:Pitch}
\end{eqnarray}
with a supercritical bifurcation at $\mu=0$. The stable stationary solutions of the system are for $\mu \le 0$, $x_s=y_s=0$, and for $\mu > 0$, $x_{s,\pm}=\pm \sqrt{2 \mu}, y_s=-\mu$. 

A third system that will be investigated is giving rise to a supercritical Hopf bifurcation,
\begin{eqnarray}
 \frac{dx}{dt} &=& \mu x-\omega y-(x^2+y^2)x
       +\sqrt{2D}\, \frac{dW_x}{dt}, \nonumber \\
 \frac{dy}{dt} &=& \omega x+\mu y-(x^2+y^2)y
       +\sqrt{2D}\,\frac{dW_y}{dt},
    \label{Eq:Hopf}
\end{eqnarray}
 also known as the Stuart-Landau system. For $\mu \le 0$, a stable stationary solution is found at $(0, 0)$. Beyond that point, a stable periodic oscillation is obtained that will rotate around $(0, 0)$ with a period of $2 \pi/\omega$. An interesting property of the Stuart-Landau system is that the asymptotic joint probability density does not depend on the rotation part associated with $\omega$, leading to an equivalent potential-like solution.

\section{Analysis of the rate of information transfer close to the bifurcation points}

In this section, the three different systems discussed above will be investigated analytically and numerically. We will first start with the saddle-node bifurcation present in the Mexican Hat, and then follow with the analyses of the Pitchfork and Hopf bifurcations.

\subsection{The Saddle-node bifurcation within the Mexican Hat}

\subsubsection{RIT from \texorpdfstring{$y$}{y} to \texorpdfstring{$x$}{x}}

The information transfer from $y$ to $x$ is written following \cite{Pires2024} as
\begin{equation}
T_{y \rightarrow x}=-E\! \left[ F_{x,\neq x} \frac{\partial \log p(y | x)}{\partial x}\right],
\end{equation}
where the total expectation is given by
\begin{equation}
E(g)=\iint dx\,dy\,p(x,y)\,g(x,y).
\end{equation}
The deterministic x- forcing term not depending on variable $x$ is
\begin{equation}
F_{x,\neq x}=-\alpha x y^2.
\end{equation}

Using the above relationship, the RIT can be written as (see Appendix \ref{Appendix2}),

\begin{eqnarray}
T_{y\to x}
&=&\frac{\alpha^2}{D}\int_{\Omega_x} dx\,x^2p(x)
\int_{\Omega_y}
\left(y^2-m_2(x)\right)^2p(y\mid x)\,dy \nonumber \\
&=&\frac{\alpha^2}{D}\int_{\Omega_x} dx\,x^2p(x)
\left[m_4(x)-m_2^2(x)\right], \nonumber \\
&=& \frac{\alpha^2}{D} E_x(x^2 Var(y^2|x))
\label{Eq:Tyx-MexHat}
\end{eqnarray}
where
\begin{equation}
m_4(x)=\int_{\Omega_y}y^4p(y| x)\,dy, \,\,\,\,\,\,\,\,\,
m_2(x)=\int_{\Omega_y}y^2 p(y| x) \,dy, \,\,\,\,\,\,\, Var(y^2|x)=m_4(x)-m_2^2(x) \nonumber
\end{equation}
and the probability density functions (PDFs) are defined as
\begin{eqnarray}
p(x,y)&=&\frac{1}{Z}\exp[-\frac{U(x,y)}{D}] \nonumber \\
p(x)&=& \frac{1}{Z}\int_{\Omega_y} dy\,e^{-U(x,y)/D} \nonumber \\
p(y\mid x)&=&\frac{p(x,y)}{p(x)}
=\frac{e^{-U(x,y)/D}}{\displaystyle\int_{\Omega_y} dy\,e^{-U(x,y)/D}}. \nonumber
\end{eqnarray}
with $Z=\iint_{\Omega} dx\,dy\,\exp[-\frac{U(x,y)}{D}]$.

Equation \ref{Eq:Tyx-MexHat} is the first important relation that will be used later in the numerical analysis for this system, which shows a dependence on the bifurcation parameter $\alpha$ and of the noise term $D$. This however provide a limited understanding of the properties of this rate of information transfer because of the implicit dependence of the joint PDF on the parameters.  In order to grasp better the role of those parameters, a small noise limit is taken that can then be analyzed while approaching the bifurcation point.

\bigskip
{\bf Small noise limit}
\bigskip

Let us now place ourselves in the case where the noise is small $D \rightarrow 0$ and below the bifurcation point at $x_s=\frac{1}{\sqrt{3}}$. The details of the computation are given in Appendix \ref{Appendix3}, and gives
\begin{equation}
T_{y\to x}
\simeq \frac{D}{2}\frac{x_s^2}{(1-x_s^2)^2}
\label{Approx-y-to-x}
\end{equation}
where $x_s$ is the stable branch given by
\begin{equation}
x_s(\alpha)=\frac{2}{\sqrt{3}}
\cos\left[
\frac13\arccos\left(\frac{3\sqrt{3}\,\varepsilon}{2\alpha}\right)
-\frac{2\pi}{3} \right]
\end{equation}
Relation \ref{Approx-y-to-x} suggests that for $x_s$ approaching the limit point at $1/\sqrt{3}$ from below, a constant value is reached. This point will be taken up when the numerical analysis is performed.  

\subsubsection{RIT from \texorpdfstring{$x$}{x} to \texorpdfstring{$y$}{y}}

For the opposite transfer from $x$ to $y$,
\begin{equation}
T_{x\to y}
= E \left(F_{y,\neq y} {\partial y}\log p(x | y)\right),
\end{equation}
with $F_{y,\neq y}=-\alpha x^2 y$, one can compute similarly with details that can be found in Appendix \ref{Appendix4}. One gets a similar formula than with the rate of information transfer $y \to x$,
\begin{equation}
T_{x\to y} =\frac{\alpha^2}{D} \int dy\,y^2p(y) \left[ m_4(y) -(m_2(y))^2 \right]
\label{Eq:Txy-MexHat}
\end{equation}
where
\begin{equation}
m_4(y)=\int_{-\infty}^{+\infty}x^4p(x| y)\,dx, \,\,\,\,\,\,\,\,\,
m_2(y)=\int_{-\infty}^{+\infty}x^2 p(x| y) \,dx. \nonumber
\end{equation}
This expression is very similar as \ref{Eq:Tyx-MexHat} but as we will see below, the asymmetry of the bifurcation in $x$ and $y$ will lead to different small noise properties, and numerical behavior close to the bifurcation.

\bigskip
{\bf Small noise limit}
\bigskip

In the small noise limit discussed in Appendix \ref{Appendix5}, one gets an expansion of the RIT as
\begin{equation}
    T_{x\to y}
\simeq \frac{4 x_s^2D}{(3x_s^2-1)(x_s^2-1) }
\label{Approx-RIT-x-to-y}
\end{equation}
 It is important to underline here that Eq. \ref{Approx-RIT-x-to-y} is now diverging when close to the limit point, $x_s=1/\sqrt{3}$ as revealed by the denominator of \ref{Approx-RIT-x-to-y}. This contrasts with the rate of information transfer in the other direction $y \to x$, but it is reminiscent of the divergence that is experienced with the variance or the autocorrelation when approaching a saddle-node bifurcation \cite{Scheffer2009, Peters2012, Kuehn2013}. 

\subsubsection{Numerical analysis}

\begin{figure}
    \centering
    \includegraphics[width=0.9\linewidth]{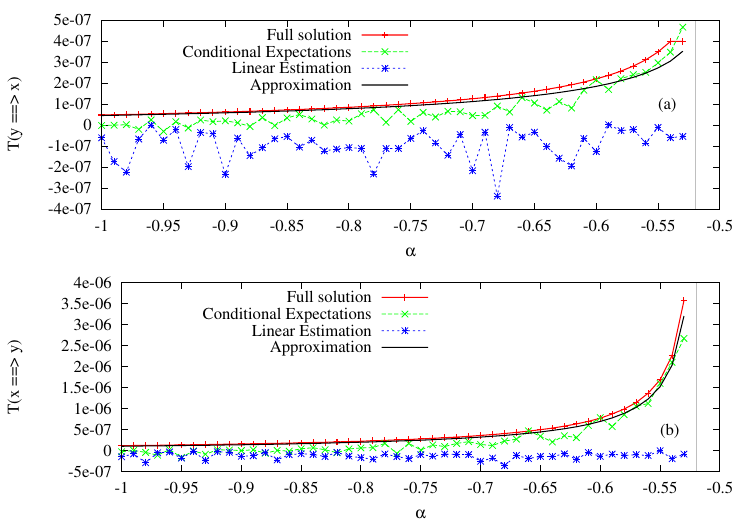}
    \caption{The rate of information tranfer for the Mexican Hat model. Panel (a) displays the rate of information transfer from $y$ to $x$ while panel (b) displays the rate of information transfer from $x$ to $y$. Four curves are displayed: The full solution using Eqs. \ref{Eq:Tyx-MexHat} or \ref{Eq:Txy-MexHat} is shown with the red curve with pluses, the approximation based on the asymptotic expansion for small noise is given by the black continuous curve without symbols; the estimation using the conditional expectations given by Eq. \ref{totalret} is provided as the green line with crosses; and finally, the blue dotted curve with stars represents the linear approximation of \cite{Liang2014}. The value of $\alpha_c=-\frac{3\sqrt{3}}{2} 0.2=-0.52$. Values are displayed for $|\alpha|>|\alpha_c|$ or $\Delta<0$ before the saddle-node bifurcation.}
    \label{Fig:RIT-MexHat}
 \end{figure}

To validate the expressions presented above, a numerical integration of the Mexican hat model is performed using the Euler-Maryama scheme with a time step of 0.01. The parameters are fixed as $\varepsilon=0.2$ and $D=5. \, 10^{-7}$. Figure \ref{Fig:RIT-MexHat} displays the rate of information transfer from $y \to x$ (a) and $x \to y$ (b) using the different expressions and approximations discussed above. These results were obtained by integrating the model for $20,000,000$ time units for each parameter values after a transient of $10,000,000$ time units. This operation is repeated 10 times, and then an average over the ten different realizations is plotted.

The results reveal that first, the asymptotic approximations are providing valid descriptions close to the bifurcation point: $T_{y \rightarrow x}$ is slowly changing while approaching the bifurcation with a saturation at a constant value near $\frac{9D}{8}$; and $T_{x \rightarrow y}$ is diverging as approaching the bifurcation. It makes of this rate of information transfer an interesting indicator of proximity to the limit point bifurcation. Second, the use of the conditional expectation expression provided a result in close agreement with the reference full solution provided by Eqs. \ref{Eq:Tyx-MexHat} and \ref{Eq:Txy-MexHat}. Finally the linear approximation clearly fails in estimating properly the transfer of information, and does not show any signal of proximity to the bifurcation. 

It is worth noting that, as the bifurcation point is approached, the potential barrier ensuring the local confinement of trajectories progressively decreases. For finite noise amplitudes, this may compromise the validity of the locally normalized probability densities used in the analytical calculations. Caution is therefore required when interpreting the analytical results in the immediate vicinity of the bifurcation point. Nevertheless, the asymmetry between the two directional rates of information transfer (RITs) remains a robust feature of the behavior observed when approaching the saddle-node bifurcation.

Interpreting the asymmetry between the two RITs is tricky. If one looks at the expression of the rate of information transfer as computed using the approach proposed in \cite{Pires2024}, one can reach a better understanding. Let us come back to the expressions of the rate of information transfer in the Eq. \ref{totalret} for the specific model under interest,
\begin{eqnarray}
    T_{y \rightarrow x} &=&  E_x \left[  -\alpha x \frac{\partial E(y^2| x)}{\partial x}  \right] \\ 
    T_{x \rightarrow y} &=& E_y \left[  -\alpha y  \frac{\partial E (x^2| y)}{\partial y}  \right]
\end{eqnarray}

The key terms here are the derivatives of the conditional expectations. When analyzing the numerical results, it is evident that the conditional expectation $E(y^2| x)$ does not depend much on $x$ when close to the bifurcation, implying that the derivative against $x$ is very small and does not diverge. On the contrary, when considering $E(x^2| y)$ the dependence is much stronger, with peaks for the extreme values of $y$. This suggest that the conditional variance of $x$ is highly dependent on $y$ in particular for extreme values of $y$, and gives rise to the divergence found analytically. 

\subsection{The Pitchfork bifurcation}

For the Pitchfork bifurcation described above, the rate of information transfer are given by,
\begin{equation}\label{eq:transfer-definition}
 T_{y\to x}=E \left[F_{x, \ne x}
   \frac{\partial\log p(y \mid x)}{\partial x} \right],\qquad
 T_{x\to y}=E\left[F_{y, \ne y}
   \frac{\partial\log p(x\mid y)}{\partial y}\right].
\end{equation}
Here $F_{x, \ne x}$ and $F_{y,\ne y}$ denote as before only the terms that do not depend on
the target variable alone. Here those terms are
\begin{equation}
     F_{x, \ne x}=-xy, \qquad  F_{y, \ne y}=- \frac{1}{2} x^2.
\end{equation}

\subsubsection{RIT from \texorpdfstring{$y$}{y} to \texorpdfstring{$x$}{x}}
\label{Tyx-Pitch}

Let us start with the joint probability density expressed in terms of the potential,
\begin{equation}\label{eq:joint}
 p_s(x,y)=\frac1{Z}
 \exp\!\left[-\frac1D \left(\frac{x^4}{8}-\frac{\mu x^2}{2}
 +\frac12\left(y+\frac{x^2}{2}\right)^2\right)\right].
\end{equation}
From there, the conditional probability density $p(y \mid x)$ and the marginal probability density can be deduced,
\begin{eqnarray}
 p(y\mid x) & = & \frac1{\sqrt{2\pi D}} e^{-(y+x^2/2)^2/(2D)}, \nonumber \\
 p(x) & = &\frac1{Z_c}  \exp\left[-\frac1D\left(\frac{x^4}{8} -\frac{\mu x^2}{2}\right)\right]  \nonumber
\end{eqnarray}
where $ Z_c = \int_{-\infty}^{\infty}
 e^{-[x^4/8-\mu x^2/2]/D}\, dx.$ Note also that $y$ is conditionally sampled from a Gaussian distribution $N(-x^2/2,D)$.

 The RIT can then be explicitly computed as demonstrated in Appendix \ref{Ap:Pitch-yx} giving,
 \begin{equation}
      T_{y\to x}= E_x[x^2]
      \label{Eq:tyx-Pitch}
 \end{equation}
 which is just the variance of $x$. This will then follow the classical law of divergence for small noise already demonstrated in \cite[e.g.][]{Kuehn2013}, namely,
 \begin{equation}\label{Eq:Tyx-sn-Pitch}
 T_{y\to x}\sim\frac{D}{|\mu|},
\end{equation}
Note however that when approaching $\mu=0$ from below, higher order terms will play a role and the value of the transfer of information will saturate at a finite value, which is $0.956 \sqrt{D}$. This aspect will be taken up in the numerical verification.

\subsubsection{RIT from $x$ to $y$}

The same procedure can be taken here, with the joint probability density already defined in the previous section \ref{Tyx-Pitch}. The details are provided in Appendix \ref{Ap:Pitch-xy}, which give the exact compact expression,
\begin{equation}
\label{Eq:txy-Pitch}
 T_{x\to y}=\frac1{4D} E_y\left[Var(x^2\mid y)\right].
\end{equation}
where $Var(x^2\mid y)=m_4(y)-m_2(y)^2$. This already shows a very different picture here as the pure variance dependence of $T_{y \to x}$. 

Now let us consider the small noise limit. In this case, $p(x|y)\ \simeq\ N \left(0,\frac{D}{y-\mu}\right)$, and 
\begin{align}
 Var(x^2\mid y)  &=E[x^4\mid y]-E[x^2\mid y]^2\\
 &\simeq 3\left(\frac{D}{y-\mu}\right)^2  -\left(\frac{D}{y-\mu}\right)^2 =\frac{2D^2}{(y-\mu)^2}.
\end{align}
Using that $\mu$ dominates over $y$ for small noise (not too close to the bifurcation point), we have 
\begin{equation}
\label{Eq:Txy-sn-Pitch}
T_{x\to y}\simeq \frac{D}{2\mu^2}. 
\end{equation}
The latter result is very interesting as it shows that the RIT changes faster from $x$ to $y$ than the other way around. 

\subsubsection{Numerical verification}

\begin{figure}
    \centering
    \includegraphics[width=0.9\linewidth]{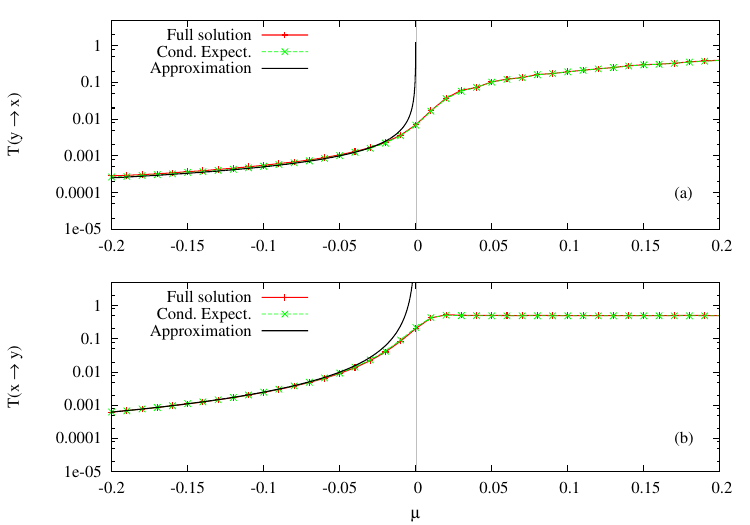}
    \caption{The rate of information tranfer for the Pitchfork bifurcation. Panel (a) displays the rate of information transfer from $y$ to $x$ while panel (b) displays the rate of information transfer from $x$ to $y$. three curves are displayed: The full solution using Eqs. \ref{Eq:tyx-Pitch} and \ref{Eq:txy-Pitch} is the red curve with pluses; the approximation based on the asymptotic expansion for small noise is given by the black continuous curve without symbols (Eqs.\ref{Eq:Tyx-sn-Pitch} and \ref{Eq:Txy-sn-Pitch}) ; and the estimation using the conditional expectations given by Eq. \ref{totalret} is provided as the green line with crosses.}
    \label{Fig:RIT-Pitch}
\end{figure}

To validate those results, we perform a numerical analysis by integrating the equations using the Euler-Maryama scheme with a time step of 0.01, with a noise term $D=5. 10^{-5}$. Figure \ref{Fig:RIT-Pitch} displays the exact estimation coming from \ref{Eq:tyx-Pitch} and \ref{Eq:txy-Pitch}. Those are the red lines with pluses symbols in panels (a) and (b). The green curve with crosses represents the estimation using the conditional expectations from \cite{Pires2024} also given at Eq. \ref{totalret}. It is interesting to see that a very good match is found between the two curves, revealing the consistency of the current analysis with the relationships derived in \cite{Pires2024}, which employs other ways to make the estimations. Finally the black continuous curve represents the approximations given in Eqs. \ref{Eq:Tyx-sn-Pitch} and \ref{Eq:Txy-sn-Pitch}. The approximation is valid up to a certain value of $\mu < 0$ and then progressively diverges close to $\mu=0$. Note that at the bifurcation point, finite values are reached.

As for the saddle-node bifurcation, an asymmetry is present with a a larger sensitivity of the RIT from the variable experiencing the bifurcation (here $x$) toward the other variable $y$. In this situation, we may consider that the variable experiencing the bifurcation is providing more uncertainty on the outcome of the other variable.

\subsection{The Hopf bifurcation}

Let us now move to the Hopf bifurcation occurring in system \ref{Eq:Hopf}. In this case, the stationary density is
\begin{equation}
 p(x,y)=\frac{1}{Z} \exp\left[ -\frac{1}{D} \left( \frac{-\mu}{2}(x^2+y^2) +\frac14(x^2+y^2)^2 \right) \right].
 \label{Eq:joint-Hopf}
\end{equation}
in which the circulation term related to $\omega$ disappears, and $\mu<0$. This feature is interesting here as it makes the system as an effective potential system, and one can make the analysis as we did for the saddle-node and Pitchfork bifurcations. Let us now compute the two RITs.

\subsubsection{RIT from \texorpdfstring{$y$}{y} to \texorpdfstring{$x$}{x}}

Let us first rearrange the terms in the effective potential appearing in \ref{Eq:joint-Hopf} as
\begin{equation}
    U_{eff} (x,y)=\left[ \frac{-\mu}{2}x^2+\frac{1}{4}x^4+\frac{x^2-\mu}{2}y^2+\frac14y^4 \right]
\end{equation}
and we can now compute the conditional probability keeping in mind that the terms only dependent on $x$ are disappearing with the normalization,
\begin{equation}
 p(y|x) = \frac{1}{Z_c} \exp\left[ -\frac{1}{D} \left( \frac{x^2-\mu}{2}y^2+\frac14y^4 \right) \right],
 \label{Eq:conditional-Hopf}
\end{equation}
where
\begin{equation}
 Z_c = \int_{-\infty}^{\infty} \exp\left[ -\frac{1}{D}
 \left( \frac{x^2-\mu}{2}y^2+\frac14y^4\right) \right]dy.
 \label{eq:Zcdef}
\end{equation}
whose estimation is left to the Appendix \ref{Ap:Norm-Tyx-Hopf},
\begin{equation}
Z_c = \sqrt{\frac{x^2-\mu}{2}}\, \exp\left(\frac{(x^2-\mu)^2}{8D}\right)
K_{1/4}\left(\frac{(x^2-\mu)^2}{8D}\right).
\label{eq:ZcBessel}
\end{equation}
where the function \(K_\nu(z)\) is the modified Bessel function of the second kind. Let us now compute the RIT,
\begin{equation}
 T_{y\to x} = E\left[ F_{x,\ne x}\, \partial_x\ln p(y|x) \right],
 \label{eq:Tdefinition}
\end{equation}
where $F_{x,\ne x}=-\omega y - x y^2$. From the exact conditional distribution,
\begin{eqnarray}
   \partial_x\ln p(y|x) &=&
 -\frac{x}{D}y^2 - \partial_x\ln Z_c \nonumber\\
 &=& -\frac{x}{D}y^2 - \frac{\partial_x Z_c}{Z_c} \nonumber \\
 &=& -\frac{x}{D} \left[y^2-m_2(x)\right]
 \label{eq:conditionalDerivative}  
\end{eqnarray}
where the second moment in $y$ is $m_2(x)=\int_{-\infty}^{\infty} \, y^2 \, p(y\mid x) \, dy$. Let us now inject that in the equation for the RIT,
 \begin{equation}
 T_{y\to x} = -\frac{1}{D} E\left[ (-\omega y - x y^2)\, x (y^2-m_2(x)) \right],
 \label{eq:Tdefinition1}
\end{equation}
which can be decomposed into two contributions, one involving the rotation term measured by $\omega$ and the other one with the nonlinearity. These two terms can be expicitely written as,
\begin{eqnarray}
  T_{y\to x} & = & \frac{\omega}{D} E\left[xy (y^2 - m_2(x))\right]
   + \frac{1}{D} E\left[x^2 y^2 (y^2-m_2(x)\right] \nonumber \\
  & = & \frac{\omega}{D} E_x\left[ x (E(y^3 \mid x)-E(y \mid x ) m_2(x)  )      \right] + \frac{1}{D} E_x \left[ x^2 (E(y^4\mid x)-m_2(x)^2)\right]
\end{eqnarray}
using the properties of conditional expectations. As $p(y \mid x) = p(-y \mid x)$, then the first and third conditional moments in y are vanishing, and the first term disappear altogether. In other word, the rotation does not contribute here, and we are left with the second term which can be written in compact form as,
\begin{equation}
     T_{y\to x} =  \frac{1}{D} E_x \left[ x^2 Var(y^2 \mid x)     \right]
     \label{Eq:Tyx-Hopf}
\end{equation}
Note that by symmetry invariance, the RIT from $x$ to $y$ is equal to the above relationship. Using the fact that the densities are vanishing at infinity and $ \int_{-\infty}^{\infty} \frac{d}{dy} \left[ y\,e^{-U(y)/D} \right]\,dy =0, $, one gets,
\begin{equation}
 Var(y^2 \mid x) = D-(x^2-\mu) m_2(x) - m_2(x)^2.   
\end{equation}

Let us now focus on the computation of $m_2(x)$. Deriving the definition of $Z_c$ in Eq. \ref{eq:Zcdef} against $x^2$, one can prove that
\begin{equation}
\frac{\partial Z_c}{\partial (x^2)} = -\frac{Z_c}{2D} m_2(x).
\end{equation}
Now using the derivative against $x^2$ of the logarithm of Eq. \ref{eq:ZcBessel} and using the previous result, one gets,
\begin{equation}
m_2(x) = \frac{x^2-\mu}{2} \left[ \frac{K_{3/4} ((x^2-\mu)^2/(8D))}{K_{1/4} ((x^2-\mu)^2/(8D))} - 1\right] 
\end{equation}

Let us now compute the marginal density. Let us first write the joint probability density in the following form $p(x,y) = \frac{1}{\mathcal Z} e^{-\Phi(x)/D} e^{-\Psi(y)/D}$ where $\Phi(x)= \frac{- \mu}{2}x^2+\frac14 x^4$ and $\Psi(y)
= \frac{x^2-\mu}{2}y^2+\frac14 y^4$. The marginal density can then be written as,
\begin{equation}
 p(x) = \frac{1}{\mathcal Z} e^{-\Phi(x)/D} Z_c(x^2-\mu).
\end{equation}
over which the RIT should be computed. 

\bigskip
{\bf Small noise limit}
\bigskip

Let us consider first the $Z_c$ term by expanding the Bessel function when D tends to 0, one gets $Z_c=\sqrt{\frac{2 \pi D}{x^2-\mu}}$, while for $m_2(x)=\frac{D}{x^2-\mu}$. If we inject this in the rate of information transfer, one gets
\begin{equation}
T_{y\to x} = \frac{2 D \sqrt{2 \pi D}}{Z} \int_{-\infty}^{+\infty} \frac{x^2}{(X^2-\mu)^2}\, \, \frac{1}{\sqrt{(X^2-\mu)}} exp\left[ -\frac{1}{D} (\frac{-\mu}{2} x^2 + \frac{1}{4} x^4)\right] dx
\end{equation}
If we further assume that the variability of $x$ is small as compared with $-\mu$, one gets
\begin{eqnarray}
    T_{y\to x} & \simeq &\frac{2 D}{\mu^2} E(x^2)  \nonumber \\
    & \simeq & \frac{2 D^2}{|\mu|^3}
\end{eqnarray}
with the second approximation considering that the variable $x$ is coming from a Gaussian with variance $D/|\mu|$. 
 
\subsubsection{Numerical verification}

\begin{figure}
    \centering
    \includegraphics[width=0.9\linewidth]{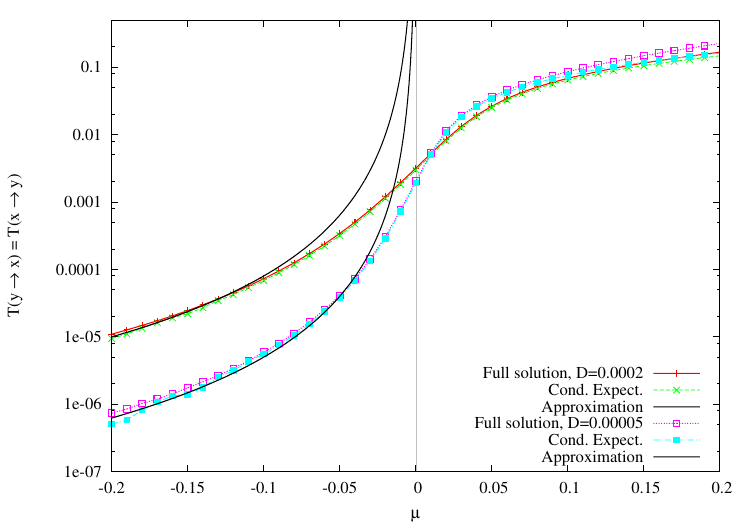}
    \caption{The rate of information transfer from $y$ to $x$ for the Hopf bifurcation for two different values of the noise amplitude $D$. three curves are displayed: The full solution using Eq. \ref{Eq:Tyx-Hopf} is the red curve with pluses and magenta with open squares, the approximation based on the asymptotic expansion for small noise is given by the black continuous curve without symbols; and the estimation using the conditional expectations given by Eq. \ref{totalret} is provided as the green line with crosses and the blue dashed curve with full squares.}
    \label{Fig:RIT-Hopf}
\end{figure}

The numerical verification has been conducted as for the other bifurcations, here with two different amplitudes of the noise $D=0.0002$ and $D=0.00005$. The results are shown on Figure \ref{Fig:RIT-Hopf}. Clearly the inverse of the cube to the proximity of the bifurcation fits well, and obviously better for the smallest value of $D$.

Interestingly, as for the Pitchfork bifurcation, a finite value is reached at the bifurcation point, which contrasts with the results obtained in the context of the saddle-node bifurcation. It is also interesting to note that the rate of change while approaching the bifurcation is much larger than for the Pitchfork bifurcation. This could bee a useful signal when approaching Hopf bifurcations.

\section{Application to low-order Climate-relevant systems}

Up to now, we have considered potential or effective potential idealized systems, that allowed us to get nice analytical approximations. Let us now focus on systems for which there is no such a potential and for which the normal form is not explicitly constructed. We will consider three systems of relevance in the climate system: The Stommel-Cessi system describing the dynamics of the thermohaline circulation over the Atlantic \cite{Stommel1961,Cessi1994}, the Jin model describing the discharge-recharge oscillation in the tropical Pacific related to the dynamics of El-Nino \cite{Jin1997}, and finally, the Jiang-Jin-Ghil model describing the large-scale dynamics of the double gyres in the ocean basins \cite{Jiang1995}.

\subsection{The Stommel-Cessi model of the Atlantic Overturning Circulation}

\begin{figure}
    \centering
    \includegraphics[width=0.7\linewidth]{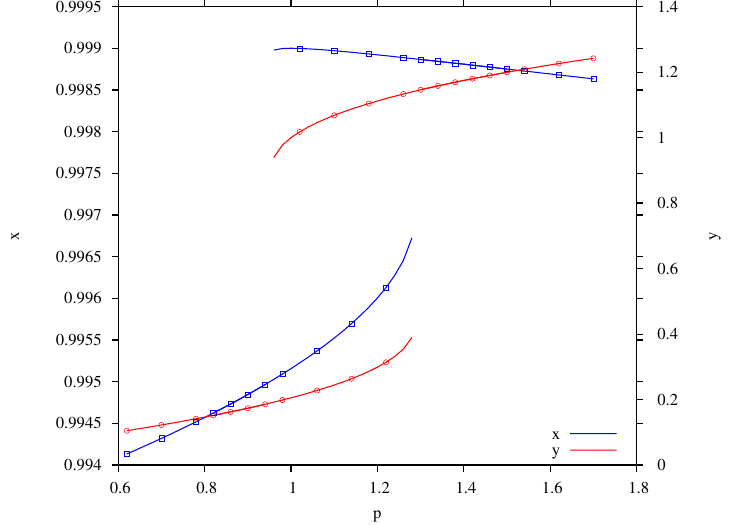}
    \caption{The bifurcation diagram showing the stable branches for $x$ (blue curve with open squares) and $y$ (red curve with open dots).It shows supercritical and subcritical saddle-node bifurcations, respectively near $p=0.95$ and $p=1.28$.}
    \label{Bifurcation-Cessi}
\end{figure}

One important characteristic of the ocean is to have a density which crucially depends on both temperature and salinity. These two elements are in fact competing in opposite ways, with an increase of temperature inducing a decrease of density, while the increase of salinity induces an increase of density. The description of the competing contributions to both salinity and temperature into the large scale dynamics of the overturning ocean circulation has led to a very famous model, known as the Stommel model \citep{Stommel1961}. This model has been revisited in the 1990s by Cessi \citep{Cessi1994} proposing different parameterizations of the large scale ocean current. One of his parameterizations is interesting as it has a structure easily amenable for our formalism leading to a simple polynomial form. This is the model which is used here,
\begin{eqnarray}
    \frac{dx}{dt} &=& -\alpha (x-1) - x (1+\mu^2 (x-y)^2) + \sqrt{2D} \frac{dW}{dt}\nonumber \\
    \frac{dy}{dt} &=& p - y (1+ \mu^2 (x-y)^2) + \sqrt{2D} \frac{dW}{dt}\nonumber \\
    & &
    \label{Cessi}
\end{eqnarray}
where $x$ and $y$ are differences of temperature and salinity between two boxes along the meridional direction in the Atlantic. The parameter $p$ is the fresh water flux which plays the role of forcing that could induce different types of stable dynamics depending on its amplitude. The parameter $\mu^2$ is the ratio between the diffusive and advective times. This system admits for some values of $p$ either a unique stable steady state, or three steady states, two of which are stable. These states are emerging through a saddle node bifurcation. Figure \ref{Bifurcation-Cessi} displays the succession of stable states that we may encounter when the parameter $p$ is increased. Each point in the diagram is the average over a long integration of Eq. \ref{Cessi} obtained for a fixed value of $p$. It mimics the catastrophic change the system may encounter when $p$ is increased, reflecting the underlying presence of the limit point bifurcation. The parameters $\alpha$ and $D$ are fixed to $1000.$ and $5. \, 10^{-7}$, respectively.

\begin{figure} 
    \centering
    \includegraphics[width=0.8\linewidth]{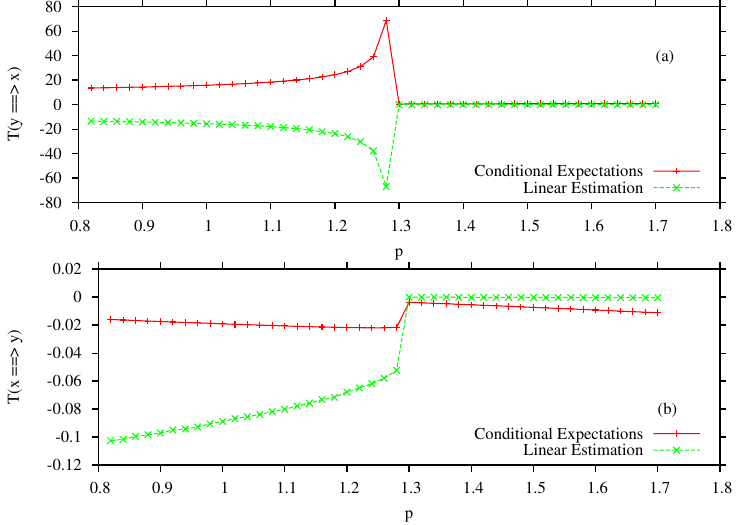}
    \caption{Rate of information transfer close to the saddle-node bifurcation present in the thermohaline circulation model of \cite{Cessi1994} when coming from the lower stable branch. Panel (a) shows the RIT from $y$ to $x$ as obtained with the conditional expectation approach of \cite{Pires2024} and with the linear approximation of \cite{Liang2014}. In panel (b) the same quantities are represented for the RIT from $x$ to $y$.}
    \label{RIT-Cessi}
\end{figure}

Following the same procedure by scanning the value of $p$ from $0.8$ to $1.7$, and letting the system stabilizing around the stationary state, the rate of information transfer can be computed. In Fig. \ref{RIT-Cessi}, the expression of the rate of information transfer based on conditional expectations given by Eq. \ref{RIT} is shown by the red curves with plus symbols for (a) $y \to x$ and (b) $x \to y$. The green curves with crosses provide the corresponding estimation using the linear approximation. 

The first remarkable result is the change of rate of information transfer $y \to x$ when approaching the saddle-node (red curve with pluses). The picture is less clear for the rate of information transfer $x \to y$ for which the rate does not vary much at the approach of the bifurcation point. This feature is clearly reminiscent of what was found for the Mexican hat with one quantity diverging while the other converging to a constant value. 

The use of the linear approximation is, also in this case, showing a drastic diverging change for the transfer from $y \to x$, but with an opposite sign. for the rate of information transfer $x \to y$, the linear approximation also shows an increase but less drastic than for the opposite transfer. Additionally, the values and sign may considerably differ from the conditional expectation estimation. This suggests that, whenever possible, it is best to use the conditional expectation estimations of \cite{Pires2024}.

\begin{figure} 
    \centering
    \includegraphics[width=0.8\linewidth]{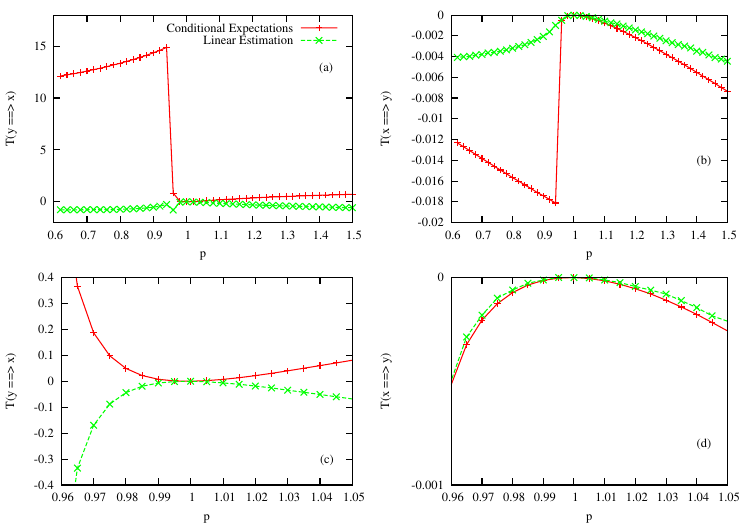}
    \caption{Rate of information transfer close to the saddle-node bifurcation present in the thermohaline circulation model of \cite{Cessi1994} when coming from the upper stable branch. Panel (a) shows the RIT from $y$ to $x$ as obtained with the conditional expectation approach of \cite{Pires2024} and with the linear approximation of \cite{Liang2014}. In panel (b) the same quantities are represented for the RIT from $x$ to $y$. Panels (c) and (d) display the same panels but reducing the range along the vertical axis in order to see the contrast of behavior between the conditional expectation and the linear estimations.}
    \label{RIT-Cessi-inv}
\end{figure}

Reversing the approach to the bifurcation point from the right to the left also shows an interesting picture as the system is moving from the upper stable branch to the lower stable branch (Fig. \ref{RIT-Cessi-inv}). Panels (a) and (b) show the overall picture with drastic changes of the RIT amplitudes. The sharp change of the RIT when approaching the bifurcation is only visible very close to the bifurcation point as revealed by the zooms at panels (c) and (d). Again the amplitudes and signs for the linear approximations may considerably differ to the conditional expectation expression Eq. \ref{totalret} of \cite{Pires2024}.

\subsection{Application to the Jiang-Jin-Ghil model}

The Jiang–Jin–Ghil model is a two-mode reduced-order representation of the wind-driven oceanic double-gyre circulation \cite{Jiang1995}.
The non-dimensional variable $x$ represents the symmetry-breaking (asymmetric) component of the circulation, while $y$ represents the symmetric double-gyre component. As the wind-stress forcing, $\mu$, increases, the symmetric state loses stability through a pitchfork bifurcation, leading to two asymmetric circulation states. The non-dimensional equations are given by
\begin{eqnarray}
\frac{dx}{dt} &=& x(y-1)\, + \sqrt{2D}\,\frac{dW}{dt},\\
\frac{dy}{dt} &=& \left(-x^2-y+\mu\right)\,
     + \sqrt{2D}\,\frac{dW}{dt}.
\label{eq:Jiang_model_stochastic}
\end{eqnarray}
where $\mu$ is the control parameter. 

\begin{figure} 
    \centering
    \includegraphics[width=0.9\linewidth]{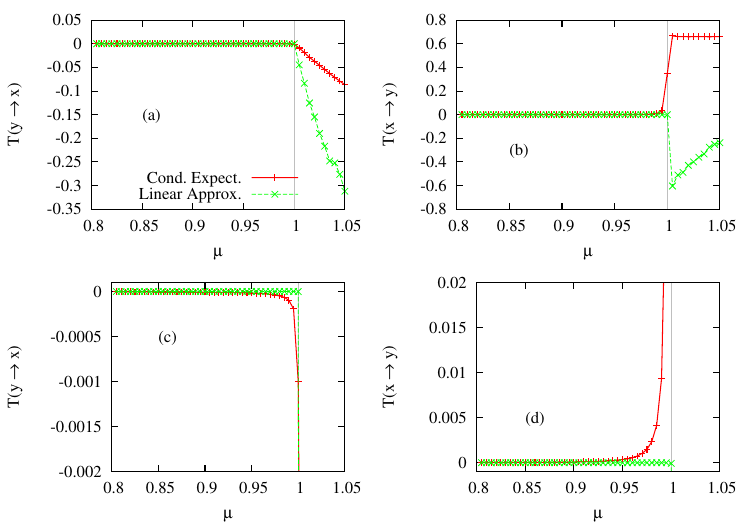}
    \caption{Rate of information transfer close to the Pitchfork bifurcation present in the double gyre model of \cite{Jiang1995}. Panel (a) shows the RIT from $y$ to $x$ as obtained with the conditional expectation approach of \cite{Pires2024} and with the linear approximation of \cite{Liang2014}. In panel (b) the same quantities are represented for the RIT from $x$ to $y$. Panels (c) and (d) display the same panels but reducing the range along the vertical axis in order to see the contrast of behavior between the conditional expectation and the linear estimations.}
    \label{RIT-Jiang}
\end{figure}

Figure \ref{RIT-Jiang} displays the rate of information transfer as obtained with the conditional expectations proposed in \cite{Pires2024} in both directions at panels (a) and (b) as a function of the control parameter $\mu$ (red curves with pluses symbols). Those overall pictures reveal a sharp change of the rate of information transfer when passing through the Pitchork bifurcation point. Note that whatever the new branches followed beyond $\mu=1$, the rate of information transfer is the same. Note also the contrast between the linear estimation and the one based on the conditional expectations. The linear approximation is indeed indicating a change but not with the right amplitude or sign.

A closer look to what is happening at the bifurcation point (see panels (c) and (d) of Fig. \ref{RIT-Jiang}) reveals that the conditional expectation estimation indeed shows a sharp increase close to $\mu=1$, suggesting that it can be used as an indicator of change. This contrasts with the linear approximation for which no early warning signal of approaching the bifurcation can be inferred. 

It is also interesting to note that the change of RIT close to bifurcation is much larger from $x$ to $y$ than in the other direction, as for the potential case discussed earlier. This consistently indicates that the information flow is changing faster from the variable experiencing the bifurcation than from the other one.

\subsection{Application to the Jin model}

The Jin's model describes the evolution of the the sea-surface temperature anomaly in the central/eastern equatorial Pacific, $x$, and of the thermocline-depth anomaly in the western equatorial Pacific, $y$. A cubic nonlinearity is introduced in the system in order to saturate the impact of the thermocline feedback on the sea surface temperature. Note that the term $\mu x$ in the nonlinear term is a parameterization of the impact of the wind stress on the ocean dynamics. See the paper \cite{Jin1997} for more information on the physics, and the coefficients present in front of the different terms. For now, we will consider the system as given,
\begin{eqnarray}
\frac{dx}{dt} &=& \left(\frac{15}{8}\mu -1\right)x +\frac{3}{4}y -\left(y+\frac{5}{2}\mu x\right)^3
+\sqrt{2D}\, \frac{dW}{dt},
\nonumber \\
\frac{dy}{dt} &=& -\frac{1}{4}y -\frac{5}{16}\mu x
+\sqrt{2D}\,\frac{dW}{dt}.
\end{eqnarray}
This system exhibits a Hopf bifurcation at $\mu=2/3$ with a new stable periodic orbit emerging from the stable stationary solution.

\begin{figure} 
    \centering
    \includegraphics[width=0.9\linewidth]{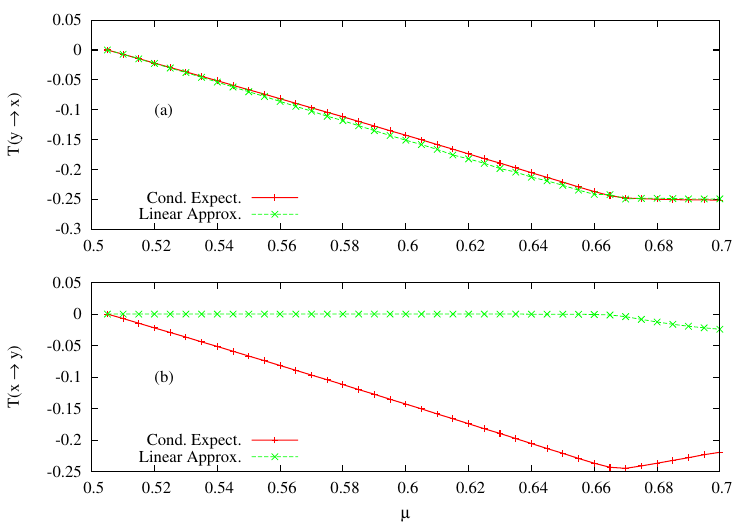}
    \caption{Rate of information transfer close to the Hopf bifurcation present in the recharge-discharge oscillator describing the dynamics of El-Nino of \cite{Jin1997}. Here $x$ stands for the sea surface temperature anomaly, and $y$ for the thermocline depth anomaly. The abscissa represents the impact of wind stress. Panel (a) shows the RIT from $y$ to $x$ as obtained with the conditional expectation approach of \cite{Pires2024} and with the linear approximation of \cite{Liang2014}. In panel (b) the same quantities are represented for the RIT from $x$ to $y$. }
    \label{RIT-Jin}
\end{figure}

Figure \ref{RIT-Jin} displays the RIT for both directions. The changes are here very smooth, contrasting with the results of the other bifurcations. It is also interesting to realize that the linear approaches provides a reasonable estimation for the RIT from $y$ to $x$ but not for the opposite direction. This reveals that the linear approximation does not necessarily identify well the impact of linear terms.

We know that a change in thermocline depth ($y$) corresponds to an upwelling/downwelling, and hence a change in the sea surface temperature (SST, $x$). This is exactly what Fig.~\ref{RIT-Jin}a reveals. However, this generally does not hold in the other way around: SST ($x$) rarely causes the thermocline depth variation ($y$), which is usually driven by wind stress near the equator or a coastline. SST may come into play only when it becomes strong enough and begins to exert influence on the wind stress above. In this sense, Fig.~\ref{RIT-Jin}b seems to be somewhat counterintuitive. More investigation is needed before a conclusion is reached.

\section{Conclusions}

The problem of estimating the rate of information transfer developed along the years by \cite{Liang2005,Liang2014,Liang2021} as been computed close to three main local bifurcations: the saddle-node, the Pitchfork and the Hopf. The analysis has been first performed within the context of potential (or quasi-potential) systems that allow for a detailed analytical derivation close to the bifurcation points, together with the small noise limit. Besides a precise information on the specific directional influence between observables close to bifurcations, the analysis reveals that 
\begin{itemize}
    \item In the case of the saddle-node bifurcation, sharp divergence of the rate of information transfer from the variable subject to the abrupt change to the second one, can be experienced. The reverse influence is also experiencing rapid changes but does not diverge at the bifurcation point.
    \item For both the Pitchfork and Hopf bifurcations, the rate of information transfers in both directions experience changes but are still finite when reaching the bifurcation point.
\end{itemize}

This analysis has been complemented by the investigation of three different nonlinear stochastic systems of climate relevance displaying such type of bifurcations, namely the Stommel-Cessi model for the thermohaline ocean circulation in the North Atlantic, the Jiang-Jin-Ghil double gyre dynamics of the ocean surface developed by \cite{Jiang1995}, and the recharge-discharge oscillator model of Jin describing the dynamics of El-Nino in the tropical Pacific. Although those non-potential models do not display the normal-form structure of the potential models, they also display drastic changes of the rate of information transfer close to the bifurcation points, which ressemble the ones found with the potential systems. An important conclusion is that the approach to the bifurcation point can be monitored using a set of causal relationships based on an ensemble of realizations of the available models, but contrasting behaviors are found depending on their nature. The sharper change are found close to the saddle-node bifurcation of the Stommel-Cessi model.

The linear approximation proposed in \cite{Liang2014} also displays such changes but the amplitudes and signs are usually very different to the ones obtained either analytically or using the approach of conditional expectations of \cite{Pires2024}, the latter providing the best estimation of the rate of information transfer for non-analytical probability distributions.

This study can be considerably extended by analyzing more sophisticated systems, either deterministic or stochastic, embedded in higher phase space dimension from low-order systems up to more realistic intermediate order models. It will allow to evaluate how those potentially sharp modifications of the rate of information transfer can survive the ladder of dimension. At the same time, the development of tools to identify such changes in higher dimensions is worth pursuing, for instance by considering the rate of information transfer from a set of variables to another set as discussed in \cite{Liang2016,Liang2021,Pires2024}. 

Another important question that should be addressed in the future is the possibility to evaluate these quantities in a real-world context, for instance for the climate system experiencing climate changes. Different strategies can be taken here. If there is enough realizations of climate predictions, averages over the set of realizations can be made to track the climate change line in \cite{Docquier2022}. In the case of the absence of such a set of parallel model integrations like for instance when dealing with the single-realization real climate evolution, very promising methods were and are currently developed to estimate locally in time the statistics involved in the computation of the rate of information transfer \cite{Hagan2019,Siddique2026}. Extensions of those methods in a fully nonlinear setting would be useful and are currently under investigations. 

\section{Declaration of generative AI and AI-assisted technologies in the manuscript preparation process}

Statement: During the preparation of this work the authors used ChatGPT (OpenAI, GPT-5.6 Sol) as an interactive aid during preliminary analytical calculations, including algebraic manipulations and the exploration of asymptotic approximations close to bifurcation points. The LLM-generated suggestions were not treated as independent scientific results. All derivations and analytical expressions reported in the manuscript were subsequently independently checked and verified by the authors, and the resulting analytical predictions were validated against numerical calculations. All the numerical analysis has been performed independently, revealing the coherence of the analytical results. After using this tool/service, the authors reviewed and edited the content as needed and take full responsibility for the content of the published article.

\section{Acknowledgments}

Carlos Pires acknowledges the support from project 2024.07832.CBM (EXCECIF) (EXploring Compound climate hazards at the European scale aiming at Causal Identification of Feedbacks),  from the Fundação para a Ciência e Tecnologia (Portugal).  The research of St\'ephane Vannitsem is partly supported by the Ministry of Education, Singapore, under its Academic Research Fund Tier 1.  X. San Liang was supported by Shanghai Institute for Mathematics and Interdisciplinary Sciences (SIMIS‐ID‐2025‐CA), and Southern Marine Science and Engineering Guangdong Laboratory (Zhuhai) (SML2023SP203).

\appendix

\section{The stationary solutions of the Mexican Hat model}
\label{Appendix1}

The stationary solutions for the Mexican Hat model satisfy
\begin{align}
(x_s^2+y_s^2-1)x_s&=\frac{\varepsilon}{\alpha}, \nonumber \\
(x_s^2+y_s^2-1)y_s&=0.
\end{align}

From the second equation, we have
\begin{equation}
y_s=0,\qquad x_s^2+y_s^2=1.
\end{equation}
As the second relation is forbidden when $\varepsilon$ is different from 0 (as we will assume in the current work), the only possible solution is $y_s=0.$

The stationary solution $x_s$ obeys therefore
\begin{equation}
x_s^3-x_s-\frac{\varepsilon}{\alpha}=0. \nonumber
\end{equation}
a depressed cubic equation that can be solved using the Cardano's method. Its discriminant is given by
\begin{equation}
\Delta=\frac{\varepsilon^2}{4\alpha^2}-\frac{1}{27}. \nonumber
\end{equation}
and the solutions are given for $\Delta<0$
\begin{equation}
x_k = \frac{2}{\sqrt3} \cos\left[ \frac13
\arccos\left( \frac{3\sqrt3\,\varepsilon}{2\alpha} \right)
-\frac{2k\pi}{3}
\right],
\qquad k=0,1,2.
\end{equation}
while only one real solution remains for $\Delta>0$  is equal to 

\begin{equation}
x_s = \frac{sgn (\frac{\varepsilon}{\alpha} )}{\sqrt3} \left( u^\frac13 + u^{-\frac13}
\right) ,  u=\sqrt {27 \Delta+1 }+\sqrt{27 \Delta} \nonumber
\end{equation}

For $\Delta=0$, two solutions converge toward a limit point (or subcritical saddle node with respect to Delta) as $|\alpha|$ decreases.
The parameter $\alpha$ at which this convergence occurs is given   by 
\begin{equation}
\frac{|\varepsilon|}{|\alpha_c|}=\frac{2}{3\sqrt{3}},
\qquad
|\alpha_c|=\frac{3\sqrt{3}}{2}|\varepsilon|. \nonumber
\end{equation}

In the numerical application, we will focus on the negative values of alpha, and the critical value is at $\alpha_c=-\frac{3\sqrt{3}}{2} \varepsilon$  at which  $x_c=\frac{1} {\sqrt{3}}$.

\section{Estimation of the RIT from \texorpdfstring{$y$}{y} to \texorpdfstring{$x$}{x} for the saddle-node bifurcation}
\label{Appendix2}

For negative values of \(\alpha\), the potential \(V(x,y)\) is not globally confining, as its quartic contribution tends to \(-\infty\) when \(x^2+y^2\to\infty\). Consequently, the formal density proportional to \(\exp[-V(x,y)/D]\) is not normalizable over the entire phase space. In the following analysis, we therefore consider a bounded region $\Omega$ containing the stable equilibrium, within which the normalizing factor can be properly normalized. This construction provides a local approximation to the probability distribution in the weak-noise regime, for observation times sufficiently short compared with the characteristic escape time from the local potential well. The restriction to a finite domain introduces boundary contributions in the analytical expressions for the rate of information transfer. These contributions are exponentially small when the potential difference between the equilibrium and the boundary is large compared with the noise intensity. The analytical results should therefore be interpreted as local weak-noise approximations. In particular, sufficiently close to the saddle-node bifurcation, the diminishing potential barrier limits the validity of these approximations, and any predicted divergence of the information transfer must be understood in this asymptotic sense.

To evaluate the RIT, the joint stationary and conditional probability densities for the potential model are
\begin{eqnarray}
p(x,y)&=&\frac{1}{Z}\exp[-\frac{U(x,y)}{D}] \\
& & \mathrm{with} \, \, Z=\iint_{\Omega} dx\,dy\,\exp[-\frac{U(x,y)}{D}] \nonumber \\
p(y)&=&\int_{\Omega_x} dx\,p(x,y)
=\frac{1}{Z}\int_{\Omega_x} dx\,e^{-U(x,y)/D} \\
p(x)&=&\int_{\Omega_y} dy\,p(x,y)
=\frac{1}{Z}\int_{\Omega_y} dy\,e^{-U(x,y)/D} \\
p(x\mid y)&=&\frac{p(x,y)}{p(y)}
=\frac{e^{-U(x,y)/D}}{\int_{\Omega_x} dx\,e^{-U(x,y)/D}} \\
p(y\mid x)&=&\frac{p(x,y)}{p(x)}
=\frac{e^{-U(x,y)/D}}{\int_{\Omega_y} dy\,e^{-U(x,y)/D}}.
\end{eqnarray}

Taking the Neperian logarithm of the conditional probability,
\begin{equation}
\log p(y\mid x)
=-\frac{\alpha}{4D}(x^2+y^2-1)^2
-\log\left[\int_{\Omega_y}
\exp\left(-\frac{\alpha}{4D}(x^2+y^2-1)^2\right)dy\right],
\end{equation}
where the factor $e^{\varepsilon x/D}$ cancels between the numerator and the normalization factor. Therefore
\begin{equation}
\frac{\partial}{\partial x}\log p(y\mid x)
=-\frac{\alpha x}{D}(x^2+y^2-1)-\frac{Z'(x)}{Z(x)}.
\end{equation}
where the conditional normalization is given by
\begin{equation}
Z(x)=\int_{\Omega_y}
\exp\left[-\frac{\alpha}{4D}(x^2+y^2-1)^2\right]dy. \nonumber
\end{equation}
and its derivative is
\begin{equation}
Z'(x)=-\frac{\alpha x}{D}
\int_{\Omega_y}(x^2+y^2-1)
\exp\left[-\frac{\alpha}{4D}(x^2+y^2-1)^2\right]dy, \nonumber
\end{equation}

The derivative of the logarithm of the conditional probability of $y$ given $x$ is then equal to
\begin{eqnarray}
\frac{\partial}{\partial x}\log p(y\mid x)
& = &-\frac{\alpha x}{D}(x^2+y^2-1)
+\frac{\alpha x}{D}\int_{\Omega_y}(x^2+y^2-1)p(y\mid x)\,dy.\nonumber \\
& = & \frac{\alpha x}{D}\left[m_2(x)-y^2\right].
\end{eqnarray}
with  $m_2(x)=\int_{\Omega_y}y^2p(y\mid x)\,dy.$ Injecting this in the equation for the rate of information transfer, one gets,
\begin{equation}
T_{y\to x} = \frac{\alpha^2}{D} \iint_{\Omega} dx\,dy\,x^2\left(y^2 -m_2(x)\right)^2p(x,y) 
\label{Final-RIT}
\end{equation}
and using $p(x,y)=p(y\mid x) p(x)$, it becomes
\begin{eqnarray}
T_{y\to x}
&=&\frac{\alpha^2}{D}\int_{\Omega_x} dx\,x^2p(x)
\int_{\Omega_y}
\left(y^2-m_2(x)\right)^2p(y\mid x)\,dy \nonumber \\
&=&\frac{\alpha^2}{D}\int_{\Omega_x} dx\,x^2p(x)
\left[m_4(x)-m_2^2(x)\right] \nonumber \\
&=& \frac{\alpha^2}{D} E_x(x^2 Var(y^2|x))
\end{eqnarray}
where $m_4(x)=\int_{\Omega_y}y^4p(y\mid x)\,dy$. This gives the formal expression of the rate of information transfer from $y$ to $x$. 

\section{The small noise limit for the RIT from \texorpdfstring{$y$}{y} to \texorpdfstring{$x$}{x}}
\label{Appendix3}

To compute the small noise limit, the potential is first developed around the $y_s=0$ as
\begin{equation}
U(x,y)=U(x,0)+\frac{1}{2} U_{yy}(x,0)y^2+\cdots. \nonumber 
\end{equation}
with $U_{yy}(x,0)=\alpha(x^2-1) > 0$, so that the conditional density is locally Gaussian at $y_s=0$,
\begin{equation}
p(y\mid x)\simeq
\sqrt{\frac{\alpha(x^2-1)}{2\pi D}}
\exp\left[-\frac{\alpha(x^2-1)}{2D}y^2\right],
\end{equation}
with a mean and variance,
\begin{equation}
\mu_y(x)=0,
\qquad
\sigma_y^2(x)=\frac{D}{\alpha(x^2-1)}.
\label{meanvar}
\end{equation}

For the marginal density in $x$, near $y_s=0$ the potential along the x-axis is given by
\begin{equation}
U(x,0)=\frac{\alpha}{4}(x^2-1)^2-\varepsilon x. \nonumber 
\end{equation}
In the case of a small amplitude noise and using the Laplace's method, the marginal density can be written as
\begin{equation}
p(x) = \frac{1}{Z_{Lapl}} \sqrt{\frac{2 \pi D}{\alpha (x^2-1)}}
\exp\left[-\frac{U(x,0)}{D}\right].
\end{equation}
where $Z_{Lapl}$ is the modified normalization factor using the Laplace's method given by
\begin{equation}
Z_{Lapl}=\frac{2\pi D}{\sqrt{\alpha^2(x_s^2-1)(3x_s^2-1)}} exp[-U(x_s,0)/D]. \nonumber
\end{equation}
Let us now assume that the potential can be expanded in Taylor series and limited to the second term,
\begin{equation}
 U(x,0) \simeq U(x_s,0)+\frac{1}{2} \frac{\partial^2 U}{\partial x^2} (x-x_s)^2  \nonumber
\end{equation}
with $U_{xx}(x_s,0)=\alpha(3x^2-1) > 0$.  Then the marginal density can be written as,
\begin{equation}
p_{Lapl}(x) \approx \sqrt{\frac{\alpha (3 x_s^2-1)}{2 \pi D}}
\exp\left[-\frac{\alpha (3 x_s^2-1)}{2D} (x-x_s)^2\right].
\end{equation}
As $m_4(x)=\mu(x)^4+6 \mu(x)^2 \sigma(x)^2+3 \sigma^4$ and $m_2(x)=\mu(x)^2+\sigma(x)^2$ where the mean and variance are given in Eqs. \ref{meanvar},
\begin{eqnarray}
    m_4(x)-m_2(x)^2&=&4\mu(x)^2 \sigma(x)^2+2\sigma(x)^4 \nonumber \\
    & = & \frac{2 D^2}{\alpha^2 (x^2-1)^2},
\end{eqnarray}
the rate of information transfer can then be written as
\begin{eqnarray}
T_{y\to x}
\simeq \frac{\alpha^2}{D}
\int_{\Omega_x} dx\,x^2
\left[\frac{2 D^2}{2\alpha^2(1-x^2)^2}
\right] \sqrt{\frac{(1-3 x_s^2)}{2 \pi D}}
\exp\left[-\frac{(1-3 x_s^2)}{2D} (x-x_s)^2\right]
\end{eqnarray}
and using again the Laplace's method assuming a well peaked Gaussian distribution, thanks to the small noise hypothesis: 
\begin{equation}
T_{y\to x}
\simeq \frac{D}{2}\frac{x_s^2}{(1-x_s^2)^2}
\label{Approx-y-to-x1}
\end{equation}
where $x_s$ is the stable branch given by $x_s=\frac{2}{\sqrt{3}}\cos\left[
\frac13\arccos\left(\frac{3\sqrt{3}\,\varepsilon}{2\alpha}\right)-\frac{2\pi}{3} \right]$.

\section{Computation of the rate of information transfer from \texorpdfstring{$x$}{x} to \texorpdfstring{$y$}{y} for the saddle-node bifurcation}
\label{Appendix4}

The opposite transfer from $x$ to $y$,
\begin{equation}
T_{x\to y}
= E \left(F_{y,\neq y} \frac{\partial}{\partial y}\log p(x | y)\right),
\end{equation}
with $F_{y,\neq y}=-\alpha x^2 y$, is now computed.

Take first the logarithm of the conditional density
\begin{equation}
\log p(x\mid y)
=-\frac{\alpha}{4D}(x^2+y^2-1)^2+\frac{\varepsilon x}{D}-\log Z(y),
\end{equation}
where $Z(y)=\int_{\Omega_x} \exp\left[-\frac{U(x,y)}{D}\right]dx$. Taking the derivative against $y$,
\begin{equation}
\frac{\partial}{\partial y}\log p(x\mid y)
=-\frac{\alpha y}{D}(x^2+y^2-1)-\frac{Z'(y)}{Z(y)},
\end{equation}
with
\begin{equation}
Z'(y)=-\frac{\alpha y}{D}\int_{\Omega_x} (x^2+y^2-1)e^{-U(x,y)/D}\,dx, \nonumber
\end{equation}
so
\begin{equation}
\frac{Z'(y)}{Z(y)}
=-\frac{\alpha y}{D}\int_{\Omega_x} (x^2+y^2-1)p(x\mid y)\,dx. \nonumber
\end{equation}

Therefore
\begin{equation}
\frac{\partial}{\partial y}\log p(x\mid y)
=-\frac{\alpha y}{D}
\left[x^2-\int_{\Omega_x} x^2p(x\mid y)\,dx\right].
\end{equation}
Introducing the conditional moment
\begin{equation}
m_2(y)=\int_{\Omega_x} x^2p(x\mid y)\,dx, \nonumber 
\end{equation}
one gets
\begin{eqnarray}
T_{x\to y}
&=&\frac{\alpha^2}{D}
\int_{\Omega_y} dy\,y^2p(y)
\left[
\int_{\Omega_x} x^4p(x\mid y)\,dx
-\left(\int_{\Omega_x} x^2p(x\mid y)\,dx\right)^2
\right]  \nonumber \\
&=& \frac{\alpha^2}{D}\int_{\Omega_y} dy\,y^2p(y)\,
Var(x^2\mid y).
\end{eqnarray}

This relation is formally equivalent to the one obtained for the transfer of information in the opposite direction, but the probability densities and moments differ. This will be made clear when looking at the small noise limit.

\section{Small noise limit for the RIT from \texorpdfstring{$x$}{x} to \texorpdfstring{$y$}{y}}
\label{Appendix5}

Let us now consider the limit for $D$ tending to 0, and focus on solutions close to the stable branch of the bifurcation diagram such that
\begin{equation}
\alpha x_s(x_s^2-1)-\varepsilon=0, \nonumber
\qquad y_s=0.
\end{equation}
But for fixed $y$, the conditional mean $x_*(y)$ is determined by
\begin{equation}
U_x(x_*(y),y)=
\alpha x_*(y)\left(x_*^2(y)+y^2-1\right)-\varepsilon=0. \nonumber
\end{equation}
The conditional density is then approximated by a Gaussian around $x_*(y)$,
\begin{equation}
p(x | y)\simeq
\sqrt{\frac{U_{xx}(x_*(y),y)}{2\pi D}}
\exp\left[-\frac{U_{xx}(x_*(y),y)}{2D}(x-x_*(y))^2\right].
\label{conditionalxy}
\end{equation}
with $U_{xx}(x_*(y),y)=\alpha (3x_*^2(y) + y^2 -1)$. The corresponding conditional variance is
\begin{equation}
\sigma_x^2(y) =\frac{D}{U_{xx}(x_*(y),y)}.
\end{equation}
with $x_*(y)=x_s-\frac{x_s}{3x_s^2-1} y^2$ evaluated in Appendix \ref{Appendix6}.

Now let us estimate $p(y)$. For fixed $y$, the Laplace approximation to the $x$ integral gives
\begin{equation}
p(y)\simeq \frac{1}{Z_{\mathrm{Lapl}}}
\sqrt{\frac{2\pi D}{U_{xx}(x_*(y),y)}}
\exp\left[-\frac{U(x_*(y),y)}{D}\right].
\end{equation}
where $Z_{Lapl}$ is the estimation of the normalization using the Laplace 's approximation.
Expanding furthermore the potential around $x_s$, 
\begin{equation}
U(x_*(y),y)
=U(x_s,0)+\frac12\alpha(x_s^2-1)y^2+\cdots, \nonumber
\end{equation} 
one gets
\begin{equation}
    p(y) \simeq \sqrt{ \frac{\alpha(x_s^2-1)} {2\pi D} } \exp\left[ -\frac{\alpha(x_s^2-1)}{2D}y^2 \right],
\end{equation}
where the variance of $y$ is now given by $D/(\alpha(x_s^2-1))$.

We can now inject all the previous results into the expression for the transfer of information $x\to y$,
\begin{equation}
T_{x\to y}
=\frac{\alpha^2}{D}
\int_{\Omega_y} dy\,y^2p(y)\, Var(x^2\mid y).
\end{equation}
If $p(x\mid y)$ is approximately Gaussian,
\begin{equation}
p(x\mid y) \sim \mathcal{N}\!\left(\mu_x(y),\sigma_x^2(y)\right),
\end{equation}
with
\begin{equation}
\mu_x(y)=x_*(y), \qquad \sigma_x^2(y)=\frac{D}{K(y)}, \nonumber
\end{equation}
where $K(y)=U_{xx}(x_*(y),y) =\alpha\left(3x_*^2(y)+y^2-1\right)$.
The variance of $x^2$ is given by
\begin{equation}
Var(x^2\mid y) =4\mu_x^2(y)\sigma_x^2(y)+2\sigma_x^4(y)
=\frac{4D\,x_*^2(y)}{K(y)}+\frac{2D^2}{K^2(y)}
\end{equation}

Substitution into the transfer gives
\begin{align}
T_{x\to y}
&=\frac{\alpha^2}{D}\int_{\Omega_y} dy\,y^2p(y)
\left[
\frac{4D\,x_*^2(y)}{K(y)}+\frac{2D^2}{K^2(y)}
\right] \nonumber \\
&=4\alpha^2\int_{\Omega_y} dy\,y^2p(y)\frac{x_*^2(y)}{K(y)}
+2\alpha^2D\int_{\Omega_y} dy\,y^2p(y)\frac{1}{K^2(y)}
\end{align}
For small noise $x_*^2(y)=x_s^2+O(y^2)$, $K(y)=K(0)+O(y^2)$ and $K(0)=\alpha(3x_s^2-1)$. Hence, to leading orders,
\begin{eqnarray}
T_{x\to y} & \simeq & \left[ \frac{4\alpha^2x_s^2}{K(0)} +\frac{2\alpha^2D}{K^2(0)}\right] \int_{\Omega_y} y^2p(y)\,dy \nonumber \\
&  \simeq & \left[ \frac{4\alpha^2x_s^2}{K(0)} +\frac{2\alpha^2D}{K^2(0)}\right]  \frac{D}{(\alpha(x_s^2-1))}
\end{eqnarray}

\section{Estimation of \texorpdfstring{$x_*(y)$}{x*(y)}}
\label{Appendix6}

Let us evaluate $x_*(y)$ appearing in \ref{conditionalxy}. Consider the expansion in $y^2$ as
\begin{equation}
    x_*(y)=x_s+\frac{dx_*}{d(y^2)} y^2 \nonumber
\end{equation}

Taking into account $U_x(x_*(y),y)=\alpha (x_* (x_*^2+y^2-1)-\varepsilon=0$,
the derivative of $x_*(y)$ against $s=y^2$ can be obtained by deriving 
\begin{equation}
U_{xx}\frac{dx_*}{ds}+U_{xs}=0, \nonumber
\end{equation}
as this is true for all $s$. Since
\begin{eqnarray}
U_{xs}=\alpha x, \nonumber \\
U_{xx}=\alpha (3 x_*^2 + s -1) \nonumber
\end{eqnarray}
evaluated at $t=0$, one gets
\begin{equation}
\left.\frac{d^2x_*}{d(y^2)}\right|_{y=0}
=-\frac{2x_s}{3x_s^2-1}, \nonumber
\end{equation}
and therefore
\begin{equation}
x_*(y)=x_s-\frac{x_s}{3x_s^2-1} y^2 
\end{equation}
to first order in $y^2$ and valid for $x_s < 1/\sqrt{3}$.

\section{The RIT from \texorpdfstring{$y$}{y} to \texorpdfstring{$x$}{x}: Pitchfork bifurcation}
\label{Ap:Pitch-yx}

The \emph{only} part of the deterministic term describing the evolution of $x$ that depends on $y$ is $F_{x, \ne x}=-xy$. Let us compute the derivative of the Neperian logarithm of the conditional probability,
\begin{equation}
\label{eq:scorexy}
 \frac{\partial}{\partial x} \log p(y\mid x)
 =-\frac{x}{D}\left(y+\frac{x^2}{2}\right). \nonumber
\end{equation}
The transfer definition \eqref{eq:transfer-definition} becomes
\begin{equation}\label{eq:tyx-start}
 T_{y \to x}=\frac1D E\left[
 x^2 y\left(y+\frac{x^2}{2}\right)\right].
\end{equation}
At fixed $x$, the Gaussian conditional density gives
\begin{equation}
 E[y \mid x]=-\frac{x^2}{2},\qquad
 E[y^2\mid x]=D+\frac{x^4}{4}. \nonumber
\end{equation}
Thus the key conditional expectation is calculated without approximation:
\begin{align}
 E\left[ y\left(y+\frac{x^2}{2}\right)\middle|x\right]
 &=E[y ^2\mid x]+\frac{x^2}{2}E[y \mid x] \nonumber \\
 &=D+\frac{x^4}{4}-\frac{x^4}{4}=D.
\end{align}
Taking the remaining expectation over $x$ in \eqref{eq:tyx-start},
\begin{equation}\label{eq:tyx-exact}
 T_{y\to x}= E_x[x^2]
 =\frac{\displaystyle\int_{-\infty}^{\infty}
 x^2e^{-[x^4/8-\mu x^2/2]/D}\, d x}
 {\displaystyle\int_{-\infty}^{\infty}
 e^{-[x^4/8-\mu x^2/2]/D}\, d x}.
\end{equation}
This equation has a very simple form, just related to the variance of $x$. Note that in the small noise limit, the resulting transfer is the traditional expression for the variance deduced for small noise, $D/|\mu|$.

\section{The RIT from \texorpdfstring{$x$}{x} to \texorpdfstring{$y$}{y}: Pitchfork bifurcation}
\label{Ap:Pitch-xy}

Only the source-dependent term in the $y$ drift contributes, $F_{y, \ne y}=-\frac12x^2$. To derive the relevant conditional density, group the terms of the potential that depend on $x$ while fixing $y$:
\begin{equation}\label{eq:pxy}
 p(x\mid y)=\frac1{Z_c}
 \exp\!\left[-\frac1D\left(\frac{x^4}{4}
                      +\frac{(y-\mu) x^2}{2}\right)\right],
\end{equation}
with $Z_c=\int_{-\infty}^{\infty} e^{-[x^4/4+A x^2/2]/D} \, d x$.

Define again $m_2(y)=E[x^2\mid Y=y]$ and $m_4(y)=E[x^4\mid y]$.  Differentiating the normalization under the integral sign gives
\begin{equation}
 \frac{\partial}{\partial y}\log Z_c(y-\mu,D)
 =-\frac{m_2(y)}{2D}, \nonumber
\end{equation}
and hence
\begin{equation}\label{eq:scoreyx}
 \frac{\partial}{\partial y} \log p(x\mid y)
 =-\frac{x^2-m_2(y)}{2D}.
\end{equation}
Inserting \eqref{eq:scoreyx} and $F_{y, \ne y}=-y^2/2$ in the
transfer definition,
\begin{align}
 T_{x\to y}
 &=\frac1{4D} E\left[x^2\bigl(x^2-m_2(y)\bigr)\right] \nonumber \\
 &=\frac1{4D} E_y\left[m_4(y)-m_2(y)^2\right].
\end{align}
or compactly,
\begin{equation}\label{eq:txy-exact}
 T_{x\to y}=\frac1{4D}
 E_y \left[Var(x^2\mid y)\right].
\end{equation}

In the small noise limit and not too close to the bifurcation point $\mu=0$, the conditional potential reduces to $\frac{y-\mu}{2}x^2$ and thus Gaussian to the leading order (with $\mu <0$):
\begin{equation}
 p(x\mid y) \simeq\  N \left(0,\frac{D}{y-\mu}\right). \nonumber
\end{equation}
The fourth-moment identity for a centered Gaussian with variance
$\sigma^2$ is $E[x^4]=3\sigma^4$, so
\begin{align}
 Var(x^2\mid y) &= E[x^4\mid y]-E[x^2\mid y]^2 \nonumber\\
 &\simeq  3\left(\frac{D}{y-\mu}\right)^2 -\left(\frac{D}{y-\mu}\right)^2 =\frac{2D^2}{(y-\mu)^2}.
\end{align}
Since the variability of $y$ is small when $D$ is small,
this becomes $2D^2/\mu^2$ at leading order after averaging. Equation \eqref{eq:txy-exact} now gives
\begin{equation}\label{eq:txy-minus}
 T_{x\to y}\sim\frac{D}{2\mu^2},
 \qquad \mu<0,\quad D/\mu^2\to 0.
\end{equation}

\section{Evaluation of the conditional normalization}
\label{Ap:Norm-Tyx-Hopf}

We now derive the normalization factor \(Z_c\) explicitly. Starting from
Eq.~\eqref{eq:Zcdef},
\begin{equation}
Z_c = \int_{-\infty}^{\infty} \exp\left[ -\frac{x^2-\mu}{2D}y^2-\frac{1}{4D}y^4 \right]dy.
\label{eq:Zc_start}
\end{equation}
Since the integrand is even,
\begin{equation}
Z_c=2\int_0^\infty \exp\left[ -\frac{1}{4D}y^4-\frac{x^2-\mu}{2D}y^2 \right]dy. 
\label{eq:Zc_even}
\end{equation}

Introduce $\beta=\frac{1}{4D}$ and $\gamma=\frac{A}{2D}$,
and define
\begin{equation}
I(\beta,\gamma) = \int_0^\infty e^{-\beta y^4-\gamma y^2}\,dy. \nonumber
\label{eq:Ibg}
\end{equation}
Then
\begin{equation}
Z_c(A)=2I(\beta,\gamma).
\label{eq:Zc_2I}
\end{equation}
Now set $u=y^2$, $y=\sqrt{u}$ and $dy=\frac{du}{2\sqrt{u}}$. Then
\begin{equation}
I(\beta,\gamma) =\frac12 \int_0^\infty u^{-1/2} e^{-\beta u^2-\gamma u}\,du. \nonumber
\label{eq:Iu}
\end{equation}
Set $t=\sqrt{2\beta}\,u$ Then $u=\frac{t}{\sqrt{2\beta}}$, $du=\frac{dt}{\sqrt{2\beta}}$ and $u^{-1/2}=(2\beta)^{1/4}t^{-1/2}$, so that $u^{-1/2}\,du=(2\beta)^{-1/4}t^{-1/2}\,dt$. In addition,
$\beta u^2=\frac{t^2}{2}$,$\gamma u=\frac{\gamma}{\sqrt{2\beta}}t$. Therefore
\begin{equation}
I(\beta,\gamma) = \frac12(2\beta)^{-1/4} \int_0^\infty t^{-1/2}
\exp\left[ -\frac{t^2}{2}-z_0t \right]dt, \nonumber
\label{eq:It}
\end{equation}
where $z_0\equiv\frac{\gamma}{\sqrt{2\beta}}$. Using
\begin{equation}
D_{-\nu}(z) = \frac{e^{-z^2/4}}{\Gamma(\nu)} \int_0^\infty t^{\nu-1}e^{-t^2/2-zt}\,dt,
\qquad
\Re\nu>0, \nonumber
\label{eq:parabolic_integral}
\end{equation}
with $\nu=\frac12$ and $\Gamma\left(\frac12\right)=\sqrt{\pi}$, it is possible to write down the equation as \citep{Glasser2012},
\begin{equation}
I(\beta,\gamma) = \frac{\sqrt{\pi}}{2} (2\beta)^{-1/4} e^{z_0^2/4} D_{-1/2}(z_0).
\label{eq:Iparabolic}
\end{equation}
The order \(-1/2\) parabolic-cylinder function is related to the modified Bessel function of the second kind by
\begin{equation}
D_{-1/2}(z) = \sqrt{\frac{z}{2\pi}}\, K_{1/4}\left(\frac{z^2}{4}\right). \nonumber
\label{eq:DminusHalfK}
\end{equation}
Substitution gives
\begin{equation}
I(\beta,\gamma) = \frac{\sqrt{\pi}}{2} (2\beta)^{-1/4}
e^{z_0^2/4} \sqrt{\frac{z_0}{2\pi}} K_{1/4}\left(\frac{z_0^2}{4}\right). \nonumber
\end{equation}
Using \(z_0=\gamma/\sqrt{2\beta}\), this simplifies to
\begin{equation}
I(\beta,\gamma) = \frac14 \sqrt{\frac{\gamma}{\beta}}\, \exp\left(\frac{\gamma^2}{8\beta}\right) K_{1/4}\left(\frac{\gamma^2}{8\beta}\right). \nonumber
\label{eq:quartic_integral}
\end{equation}
Since \(Z_c=2I\),
\begin{equation}
Z_c(A) = \frac12 \sqrt{\frac{\gamma}{\beta}}\, \exp\left(\frac{\gamma^2}{8\beta}\right) K_{1/4}\left(\frac{\gamma^2}{8\beta}\right).
\label{eq:Zc_before_substitution}
\end{equation}

Now substitute $\beta=\frac{1}{4D}$ and $\gamma=\frac{A}{2D}$, we obtain
\begin{equation}
Z_c = \sqrt{\frac{x^2-\mu}{2}}\, \exp\left(\frac{(x^2-\mu)^2}{8D}\right)
K_{1/4}\left(\frac{(x^2-\mu)^2}{8D}\right).
\end{equation}
where the function \(K_\nu(z)\) appearing above is the modified Bessel function of the second kind.

\bibliographystyle{rss}
\bibliography{main}

\end{document}